\documentclass[conference]{IEEEtran}

\ifCLASSINFOpdf

\else

\fi
\usepackage{graphicx}
\usepackage{multirow}
\usepackage{amsmath}
\usepackage{color}
 \newtheorem{definition}{Definition}

\begin{document}
%
\title{Sense, Model and Identify the Load Signatures of HVAC Systems in Metro Stations }

\author{\IEEEauthorblockN{Yongcai Wang, \emph{Member IEEE}}
\IEEEauthorblockA{Institute for Interdisciplinary Information Sciences (IIIS)\\
Tsinghua University, Beijing, 
P. R. China, 100084\\
 wangyc@tsinghua.edu.cn}
\and
\IEEEauthorblockN{Haoran Feng}
\IEEEauthorblockA{Software School, Peking University, \\ Beijing, P. R. China, 100084\\
13810091531@163.com}
}
\author{\IEEEauthorblockN{Yongcai Wang, \emph{Member IEEE}}
\IEEEauthorblockA{Institute for Interdisciplinary Information \\ Sciences (IIIS)
Tsinghua University, \\ Beijing,  P. R. China, 100084\\
 wangyc@tsinghua.edu.cn}
\and
\IEEEauthorblockN{Haoran Feng}
\IEEEauthorblockA{National Engineering Research Center \\ of Software Engineering, Peking University, \\ Beijing, P. R. China, 100084\\
haoran@pku.edu.cn}
\and
\IEEEauthorblockN{Xiangyu Xi}
\IEEEauthorblockA{Department of Automation\\
Tsinghua University\\ Beijing, P. R. China, 100084\\
xixy10@gmail.com}
}

\maketitle

\begin{abstract}
The HVAC systems in subway stations are energy consuming giants, each of which may consume over 10, 000 Kilowatts per day for cooling and ventilation.  To save energy for the HVAC systems, it is critically important to firstly know the ``load signatures'' of the HVAC system, i.e., the quantity of heat imported from the outdoor environments and by the passengers respectively in different periods of a day, which will significantly benefit the design of control policies. In this paper, we present a novel sensing and learning approach to identify the load signature of the HVAC system in the subway stations. In particular, sensors and smart meters were deployed to monitor the indoor, outdoor temperatures, and the energy consumptions of the HVAC system in real-time. The number of passengers was counted by the ticket checking system. At the same time, the cooling supply provided by the HVAC system was inferred via the energy consumption logs of the HVAC system. Since the indoor temperature variations are driven by the difference of the \emph{loads} and the \emph{cooling supply}, linear regression model was proposed for the load signature, whose coefficients are derived via a proposed algorithm .  We collected real sensing data and energy log data from HaiDianHuangZhuang Subway station, which is in line 4 of Beijing from the duration of July 2012 to Sept. 2012. The data was used to evaluate the coefficients of the regression model. The experiment results show typical variation signatures of the loads from the passengers and from the outdoor environments respectively, which provide important contexts for smart control policies.    
\end{abstract}

\IEEEpeerreviewmaketitle

\section{Introduction}
Being backbone of transportation network, the subways are also major energy consumers. As stated in a site survey conducted in  \cite{_beijing_2013}\cite{lu_analysis_2011}, a subway line (for example the line 1 in Beijing) can consume nearly 500 thousands $kW\cdot H$ power per day in the summer season, among which, more than 40\% of energy was consumed by the Heating Ventilation and Air Conditioning (HVAC) subsystems for cooling and ventilation.  If it is possible to save the energy consumption of the HVAC system a few percents, for example 10\%, dramatical energy (nearly 20 thousands $kW\cdot H$ per line, per day) can be saved. 

A major way to save energy for the HVAC systems in established subways is to design optimal control strategies to minimize the overall energy consumption or operating cost of the HVAC systems while still maintaining the satisfied indoor thermal comfort and healthy environment\cite{wang_supervisory_2008}. To optimize the design and operation of HVAC, understanding the signature of the heating or cooling load is the critically first step, which is to estimate the the quantity of heat (or cold) imported from environments or passengers into the subway station in each time unit.  By learning the knowledge of the load signature, the HVAC system can be optimally controlled to supply only necessary cooling (or heating) efforts to meet the predicted demands, which on one hand maintains the indoor comfort, on the other hand optimizes energy consumption. 


However, because the outdoor environments and the passenger flows entering or leaving a subway station are highly dynamic, the heating (cooling) loads of a subway station are diverse and are hard to estimate.  Some existing load estimation methods for buildings use the construction details and material features to estimate the heating (cooling) loads according to different outdoor temperature, using empirical models of heat conduction, radiation and convection \cite{yan2008dest}. However these models cannot capture the special features of the subway station: i) the impacts of passenger flows; ii) the piston wind pushed by trains in tunnels ; iii) 
the complex materials and underground construction structures.  Lacking effective methods to predict the load in the subway station, current HVAC systems generally controlled by simple time-driven rules, or in passive responding mode. As a result, the mismatching of load and supply is the main reason for energy waste in current subway HVAC systems. 

To characterize the load signature of HVAC system in metro stations, in this paper, we exploit the advantages of sensing and learning technologies. In a subway station, i.e., HaiDianHuangZhuang station of line 4 of Beijing subway, we deployed different kinds of environment sensors to monitor the indoor/outdoor temperatures, humidity and CO2 moisture in realtime. The passenger flow is recorded by the ticket checking system, and the energy consumptions of the refrigerators, ventilators and cooling towers of the HVAC system are monitored by the deployed smart meters. By thermal principles, we model the load of the subway station by a regression model of the sensor readings. On the other hand, the cooling supply generated by the HVAC system is inferred by the working states and energy consumptions of the HVAC system. Since the indoor temperature variations are driven by the difference of load and cooling supply, linear equations are set up, and we proposed a search algorithm to minimize the difference between integrated load and supply for dealing with the noises of sensor measurement.

We show by the identified load signatures from the real data of HaiDianHuangZhuang station, that the load of HVAC systems in the subway stations have significant characteristics, which is jointly impacted by the outdoor temperature and the passenger flow, both of which can be predicted in working days and the weekends. Therefore, the derived load signature will be useful context for further design of optimal control strategies.

The remainder of this paper is organized as following. Related works are introduced in Section 2.  Sensor deployments and the field study in HaiDianHuangZhuang station are introduced in  Section 3.  We proposed load and supply models in Section 4.  Solution method and experimental results and verification of load signatures are introduced in Section 5.  Conclusion and further works are discussed in Section 6. 
\section{Related Work}
The autonomous, optimal control for HVAC systems has attracted great research attentions in the studies of smart and sustainable buildings \cite{kelman_analysis_2011}, which is to determine the optimal solutions (operation mode and setpoints) that minimize overall energy consumption or operating cost while still maintaining the satisfied indoor thermal comfort and healthy environment \cite{wang_supervisory_2008}. 

This goal is the same in the subway HVAC control systems.  Because the HVAC systems contain different types of subsystems, such as gas-side and water-side subsystems, the optimal control problems of HVAC are extremely difficult.  One of the difficulties is the lack of an exact model to describe the internal relationships among different components. A dynamic model of an HVAC system for control analysis was presented in \cite{tashtoush_dynamic_2005}. The authors proposed to use Ziegler-Nichols rule to tune the parameters to optimize PID controlle. A metaheuristic simulation–EP (evolutionary programming) coupling approach was developed in \cite{fong_hvac_2006}, which proposed evolutionary programming to handle the discrete, non-linear and highly constrained optimization problems.  Multi agent-based simulation models were studied in \cite{andrews_designing_2011} to investigate the performance of HVAC system when occupants are participating. In \cite{yang_optimal_2012},  swarm intelligence was utilized to determine the control policy of each equipment in the HVAC system. 

One of the most closely related work is the SEAM4US (Sustainable Energy mAnageMent for Underground Stations) project established in 2011 in Europe\cite{seam4us}. It studies the metro station energy saving mainly from the modeling and controlling aspect. Multi-agent and hybrid models were proposed to model the complex interactions of energy consumption in the underground subways\cite{serban_common_2012,roberta_ansuini_hybrid_2012}. Adaptive and predictive control schemes were also proposed for controlling ventilation subsystems to save energy \cite{giretti_energy_2012}.  

Another related work reported the factors affecting the range of heat transfer in subways \cite{hu_numerical_2008}.  They show by numerical analysis that how the heat is transferred in tunnels and stations. Reference \cite{awad_environmental_2002} studied the environmental characters in the subway metro stations in Cairo, Egypt, which showed the different environment characters in the tunnel and on the surface. 
The most related work is \cite{lu_analysis_2011}, which surveyed the energy consumption of Beijing subway lines in 2008. 

Different from these existing work, we deployed sensors and presented models to study the the load signatures and distinct features of  energy consumptionof subway HVAC systems. 
 

\section{Monitor the Thermal Dynamics in Subway Station}
\subsection{Notations}
Before introducing the deployment of sensors, we firstly define notations which will be used in this paper, which are listed in Table \ref{notations}. 
 \begin{table}[htdp]
\begin{center}
\label{notations}
\caption{Notations defined for the load and supply models}
\begin{tabular}{l|l}
\hline
Notations & Definitions\\
\hline
$L(t)$ & the quantity of thermal imported from outside to inside at $t$. \\
$T(t)$ & the indoor temperature at $t$. \\
$T_{o}(t)$ & the outdoor temperature at $t$ \\
$R_{eq}$ & heat transferring resistance from outside to inside. \\
$M_{air}$ & the volume of outdoor air input into the subway station\\
$c$ & the heat capacity of per cube air. \\
$T_{p}$ & the body temperature of people. \\
$n(t)$ & the the number of passengers at time $t$. \\
\hline
$M_{mix}$ & volume of  mixed air \\
$M_{new}$ & volume of new air \\
$M_{ac}$ & volume of cooling air \\
$\alpha$ & the proportion of new air in the mixed air.\\
$T_{ac}$ & temperature of cooling air at the outlet of refrigerator.\\
$T_{mix}$ & temperature of the mixed air. \\
$e_{ac}$ & efficiency of of the cooling air transportation. \\
$M_{z}$ & the volume of air inside the subway station.\\
\hline
\end{tabular}
\end{center}
\end{table}%
\vspace{-0.5cm}
\subsection{Sensor Deployment}
A way to capture the thermal  and environment dynamics in the subway station is to deploy sensors to measure the indoor, outdoor temperatures, passenger flows and power consumptions of the HVAC systems in real-time. In HaiDianHuangZhuang subway station, which is a transferring station between line 10 and line 4 in Beijing subway, we deployed different kinds of sensors and smart meters to measure above information. The sensors were mainly deployed in the section of line 4, which is operated by Hongkong MTR. 


We installed temperature  sensors at four points inside the subway station and two points outside the subway to monitor the indoor and outdoor temperatures $T(t)$ and $T_{o}(t)$ respectively.  Note that $T(t)$ is calculated by the average of indoor temperature sensors, so as $T_{o}(t)$.  CO2 sensors are installed inside the subway to measure the indoor air quality. The passenger flow is recorded by the ticket checking system, which is denoted by $n(t)$. Note that $n(t)$ is calculated by the sum of the checked-in and checked-out passengers from $t-1$ to $t$. 

To monitor the working state of the HVAC system, temperature sensors were installed at the inlets  and  the outlets of the refrigerators to measure the temperature of the return air $T(t)$ and the cooling air $T_{ac}(t)$.   Temperature sensors are also installed at the new air pipes and mixed air  pipes of the ventilator to measure the temperatures of new air $T_{o}(t)$ and mixed air $T_{mix}(t)$.  Note that the mixed air is the mixing  of return air and new air.  The energy consumptions of different components of the HVAC system, i.e, refrigerator, ventilator, water tower, pumps, fans etc are measured in real-time by the embedded power meters of the HVAC system.

\subsection{Observed Passenger Flow Pattern}
\begin{figure}[htbp]
\begin{center}
\includegraphics[width=2.5in]{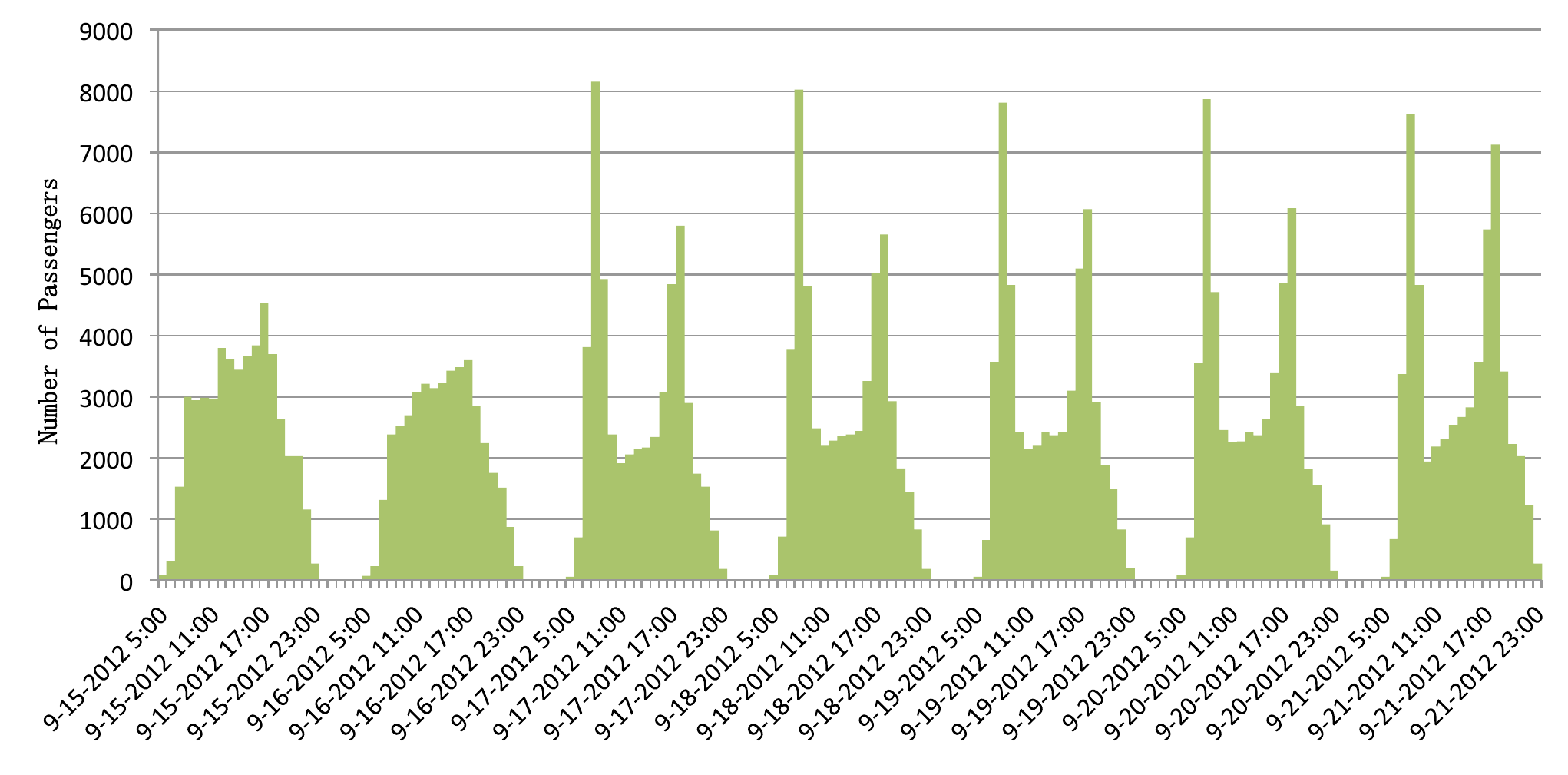}
\caption{Pattern of passenger flow over a week. }
\label{passenger}
\end{center}
\end{figure}

From the data of ticket checking system, Fig. \ref{passenger} shows the variation of passenger flow as a function of time during a week from Sep. 15 to Sep. 21. The passenger flow shows different structure in working days and weekends. In working days there are two obvious peaks in the rush hours in the morning and in the evening. In week ends, the passenger flow was almost uniformly distributed from 8:00 AM to 8:00 PM. 
\vspace{-0.5cm}
\subsection{Observed Load Signatures}
\begin{figure}[htbp]
\begin{center}
\includegraphics[width=3in]{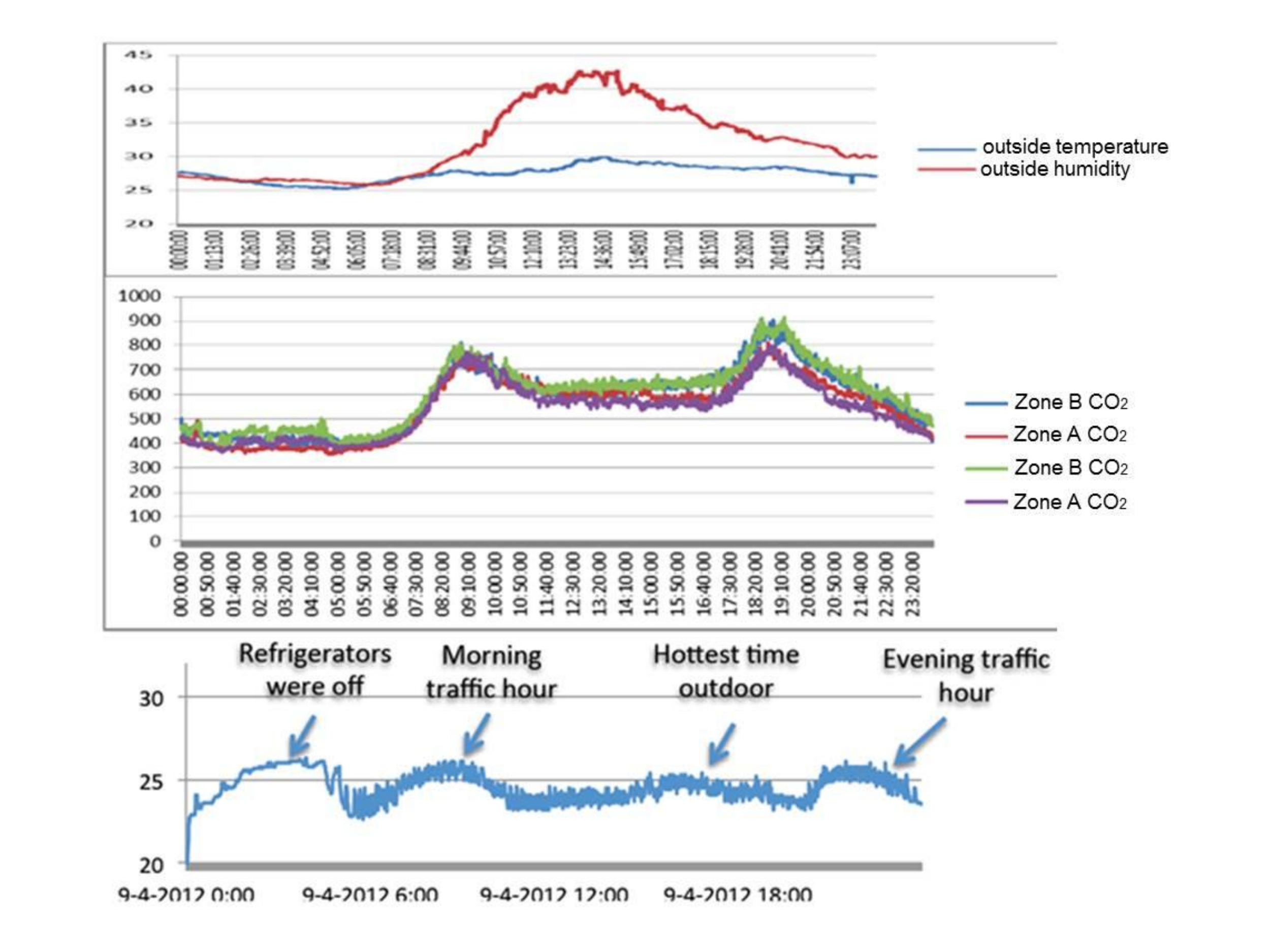}
\caption{How the indoor temperature was  affected by the outdoor temperature and passenger flow when the HVAC system was running.}
\label{signature}
\end{center}
\end{figure}

The indoor thermal condition is mainly affected by three factors: i) the outdoor temperature; ii) the passenger flow; iii) the working state of the HVAC system.  To investigate how these factors affect the indoor temperature, for a particular day, Sept. 4, February, a sunny day in 2012,  
we monitored the variations of outdoor temperature, indoor CO2 density and indoor temperature and plotted the results in Fig. \ref{signature}. It  intuitively shows how the outdoor temperatures and passenger flows affect the variation of indoor temperature. Note that during the monitoring, the HVAC system was working.

Fig. \ref{signature}(a) shows the outdoor temperature in that day. Fig. \ref{signature}(b) shows the traffic flow variation which was recorded by the ticket checking system. Fig. \ref{signature}(c ) shows the variation process of indoor temperature. By comparing these three figures, we can see that: i) the indoor temperature curve varied between 22 centigrade and 27 centigrade, which was jointly impacted by the outdoor environments, the passenger flows and the HVAC system; ii) there are four peaks in the temperature curve, which are according to following reasons:   
\begin{itemize}
\item The first peak is at 4:00 AM, which is because  the HVAC system was off in the morning, so the indoor temperature increases slowly. 
\item The second peak is at 8:00 AM, the rush hour in the morning.  It is because the quantity of thermal brought in by the passenger flow was more than the cooling effects of the HVAC system. 
\item  The third peak is at 2:00 PM, which is the hottest time in the day. This peak is not obvious, because the outdoor temperature increases slowly, the HVAC system had enough time to cool down the indoor temperature. 
\item The last peak is at 18:00 PM, the rush hour in the evening, because the cooling effects of HVAC is less than the thermal brought in by the passengers. 
\end{itemize}

These measurements show intuitively the impacts of environments and passengers to the indoor temperature.  
However a quantitative model to more accurately characterize these  impacts is still lacked. We call it the load signature, which will be modeled and learned in the next section.

\section{Model and Identify the Load Signatures}
From the sensor readings, we see the typical features of the outdoor temperature and passenger flows, but it is still unclear whose influence is more significant to the indoor temperature. In this section, we present linear regression model to identify the load signature. 
\subsection{Load Model}
\begin{definition}[load model] We define the quantity of heat imported from outdoor environments and the passengers into the subway station in a time unit as the \emph{load} of the HVAC system in the subway station. 
\begin{equation}
L\left( t \right) = \frac{{{T_o}(t)\!\!-\!\!T(t)}}{{{R_{eq}}}} + n(t)\left( {{T_p}\!\!-\!\!T\left( t \right)} \right) + c{M_{air}}\left( {{T_o}(t) \!\!-\!\!T(t)} \right)
\label{loadmodel}
\end{equation}
%
\end{definition}
$L(t)$ contains three parts:  1) the heat imported from outdoor environments by heat conduction through walls, roofs etc, i.e., $\frac{T_{o}(t)-T(t)}{R_{eq}}$; 2) the heat imported by passengers, i.e, $n(t)\left(T_{p}-T(t)\right)$; 3) the heat imported via outdoor air, i.e., $cM_{air}\left(T_{o}(t)-T(t)\right)$ which is due to piston wind or wind entered through doors. We can rewrite the equation (\ref{loadmodel}) as:
\begin{equation}
\begin{array}{l}
L\left( t \right) = c_{p}n(t)\left( {{T_p} - T\left( t \right)} \right) + \left( {c{M_{air}} + \frac{1}{{{R_{eq}}}}} \right)\left( {{T_o}(t) - T(t)} \right)\\
 = {L_p}(t) + {L_a}(t)
\end{array}
\label{load}
\end{equation}
where $L_{p}(t)=c_{p}n(t)\left( {{T_p} - T\left( t \right)} \right)$ is only related to the passengers, called the \emph{passenger introduced load (PIL)}; $L_{e}(t)=\left( {c{M_{air}} + \frac{1}{{{R_{eq}}}}} \right)\left( {{T_o}(t) - T(t)} \right)$ is caused by the indoor-outdoor temperature difference, which is called \emph{Environment Introduced Loads (EIL)}.  Note that in (\ref{load}), $T_{o}(t), T(t), n(t)$ are measured in realtime; $T_{p}$, $c$ are known constants;  only $\{c_{p}, M_{air}, R_{eq}\}$ are unknown variables.

\subsection{Supply Model}
The HVAC system runs adaptively to response the dynamics of the loads to control the indoor temperature at desired temperature. 
By assuming the indoor air is fully mixed, the variation of indoor temperature is mainly caused by the thermal difference of the load and the supply:  
\begin{equation}
L(t)-S(t)=cM_{z}\Delta(t)\label{equal}
\end{equation}
where $M_{z}$ is the volume of air in the subway station, which can be calculated by the geometrical information of the station, such as the length, width, height of the station and the tunnels. $\Delta(t)=\left(T(t+1)-T(t)\right)$
is the temperature difference changed from time $t$ to time $t+1$. 

Since the working states of the HVAC system were fully monitored, the cooling supply can be inferred by the sensors readings of the HVAC system. The HVAC system in subway station has three working modes: \begin{enumerate}
\item \emph{New air mode:}, which is used when the outdoor temperature is lower than the indoor temperature. In this mode, the refrigerator is off; The new air is the source to cool the indoor air. 
\item \emph{Refrigerator mode:} is used when the outdoor temperature is higher than the indoor temperature, during which the new air intaking  is closed and the refrigerators are working to cool the indoor air. 
\item \emph{Mixed mode:} is used when the new air's capacity is not enough to cool the indoor temperature, so both the new air ventilator and  a part of the refrigerator are working.\end{enumerate}


\begin{definition}[supply model]
We define the quantity of heat cooled down by the HVAC system in a unit time as the \emph{supply} of the HVAC system, which is defined based on different working modes of the HVAC system:
\begin{equation}
\begin{gathered}
  S(t) =  \hfill \\
  \left\{ {\begin{array}{*{20}{l}}
  {c{M_{new}}\left( {T(t) - {T_o}(t)} \right),{\textrm{              New air mode}}} \\ 
  {\left( {T_{in}^w(t) - T_{out}^w(t)} \right)V_{cool}^w\beta_{ac}{\textrm{,             Refrigerator mode}}} \\ 
  {c{M_{new}}\left( {T(t)\!\!-\!\!{T_o}(t)} \right)\!\!+\!\!\left( {T_{in}^w(t)\!\!-\!\!T_{out}^w(t)} \right)V_{cool}^w\beta_{ac}{\textrm{, Mixed}}} 
\end{array}} \right. \hfill \\ 
\end{gathered} 
\label{supply}
\end{equation}
\end{definition}


Note that in (\ref{supply}), $M_{new}$ is the volume of new air blowed into the subway station by the new air ventilator. $T_{in}^{w}(t)-T_{in}^{w}(t)$  is the temperature difference of input and output water at the refrigerator; $V_{cool}^{w}$ is the volume of the cooling water; $\beta_{ac}=c_{cool}^{w}e_{ac}$, where $c_{cool}^{w}$ is the heat capacity of the cooling water and $e_{ac}$ is the heat transportation efficiency of the refrigerator. So that $ \left( {T_{in}^w(t) - T_{out}^w(t)} \right)V_{cool}^w\beta_{ac}$ measures the cooling supply provided by the refrigerator and $cM_{new}(T(t)-T_{o}(t))$ measures the cooling supply of the new air. 


%
%

Note that $T_{o}(t), T(t), T_{in}^{w}(t), T_{out}^{w}(t)$, and $V_{cool}^{w}$ are measured in real time by the deployed sensors. $c$ is a known constant. Only $M_{new}$ and $\beta_{ac}$ are unknown. But the volume of air blowed by the ventilator in a time unit can be further inferred by the power meter readings of the ventilators. From the fan affinity laws\cite {ford2011affinity},  ventilators operates under a predictable law that the air volume delivered by a ventilator is in the one-third order of its operating power.
\begin{equation}
M_{v}=\beta_{v} E_{v}^{\frac{1}{3}}
\label{fan}
\end{equation} 
So that, the supply model of the HVAC system in the subway station can be rewritten into:  
\begin{equation}
\begin{gathered}
  S(t) =  \hfill \\
  \left\{ {\begin{array}{*{20}{l}}
  {cE_{v}^{\frac{1}{3}}\beta_{v}\left( {T(t) - {T_o}(t)} \right),{\textrm{              New air mode}}} \\ 
  {\left( {T_{in}^w(t) - T_{out}^w(t)} \right)V_{cool}^w\beta_{ac}{\textrm{,             Refrigerator mode}}} \\ 
  {cE_{v}^{\frac{1}{3}}\beta_{v}\left( {T(t)\!\!-\!\!{T_o}(t)} \right)\!\!+\!\!\left( {T_{in}^w(t)\!\!-\!\!T_{out}^w(t)} \right)V_{cool}^w\beta_{ac}{\textrm{, Mixed}}} 
\end{array}} \right. \hfill \\ 
\end{gathered} 
\label{supply1}
\end{equation}
By substituting (\ref{supply1}) and (\ref{load}), we can set up linear equations to identify the unknown parameters in the load and supply functions.  
\begin{figure*}[t]
\centering
\begin{minipage}{0.33\linewidth}
\centering
\includegraphics[width=1.8in]{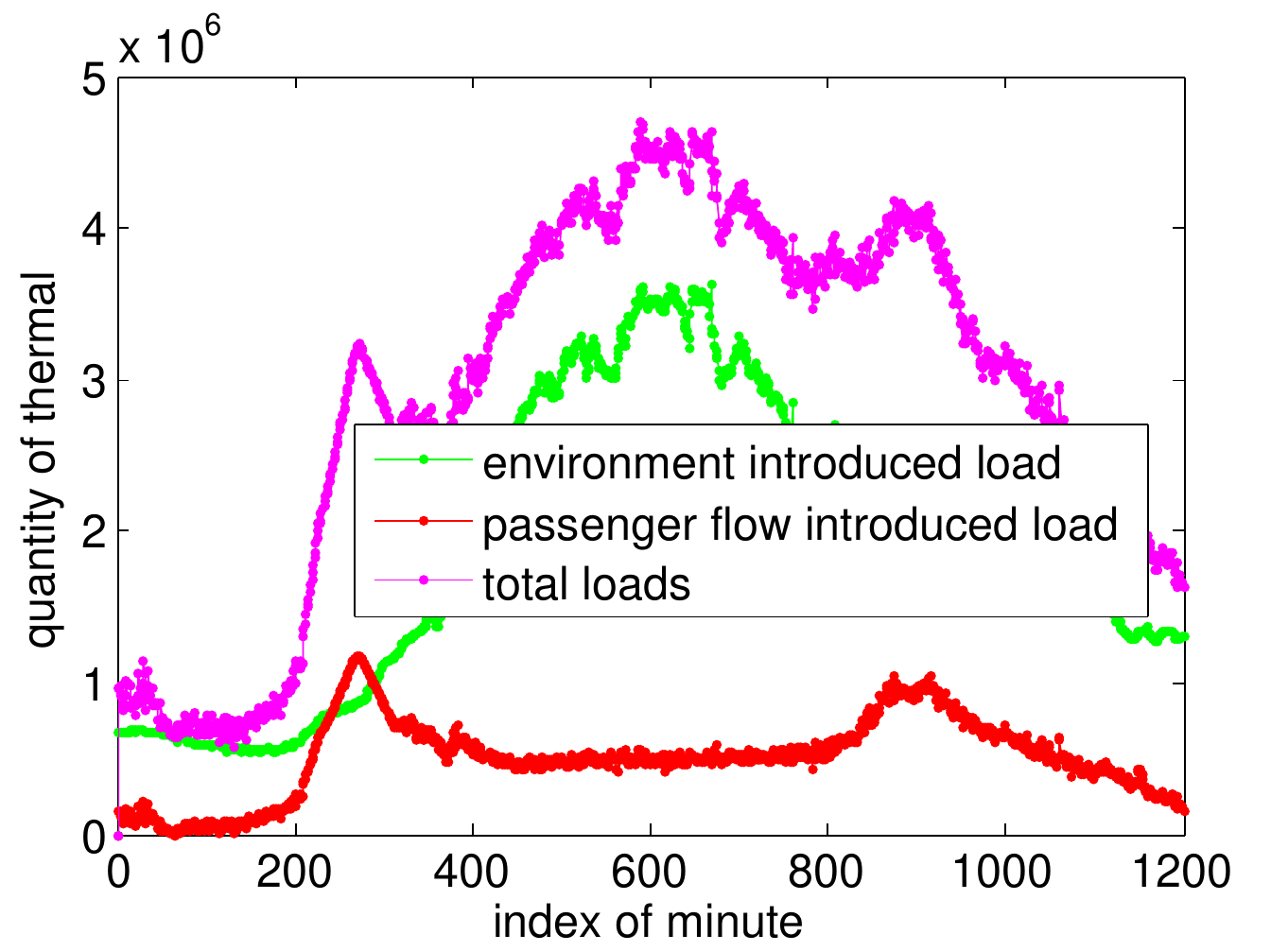}
\end{minipage}%
\begin{minipage}{0.33\linewidth}
\centering
\includegraphics[width=1.8in]{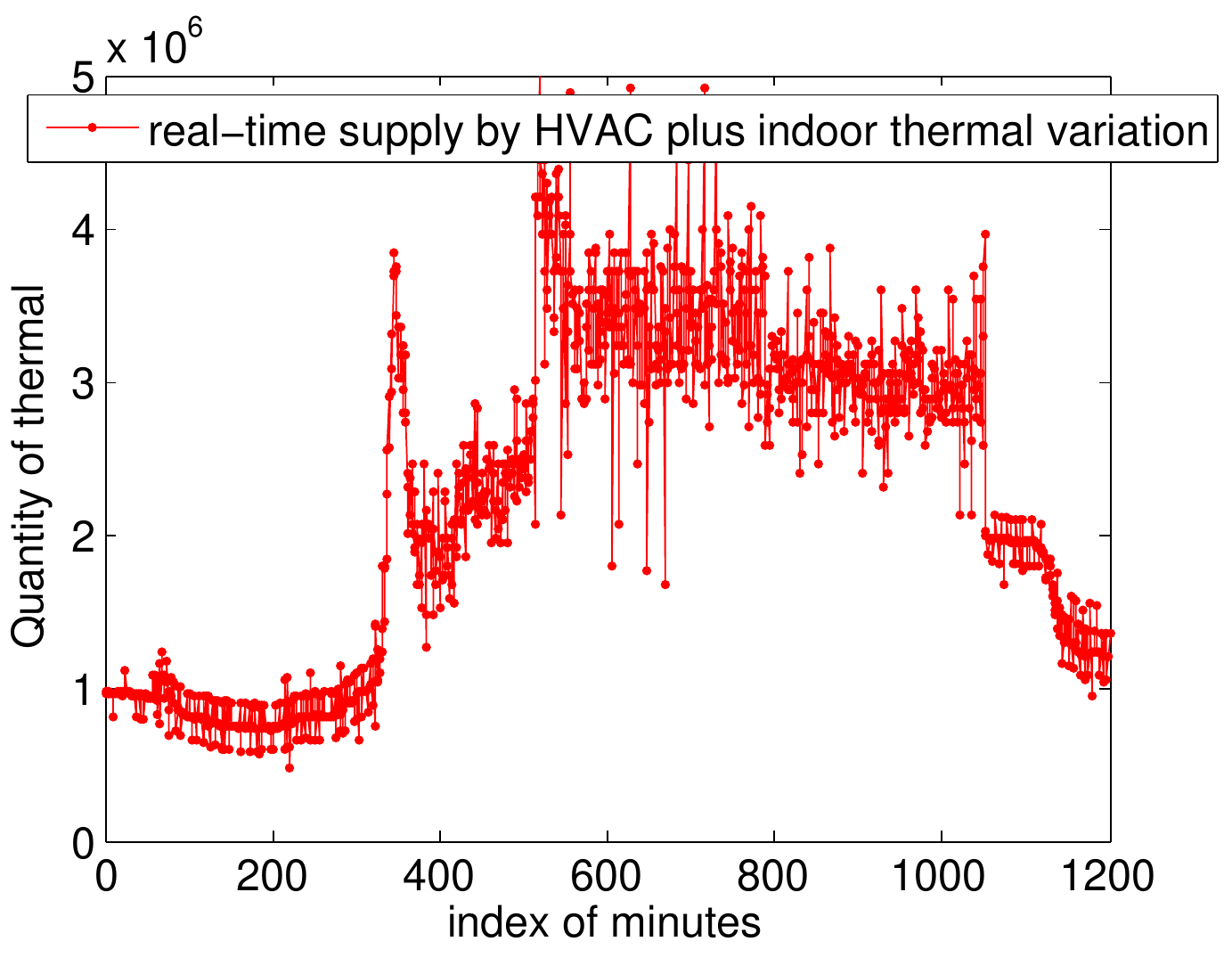}
\end{minipage}%
\begin{minipage}{0.33\linewidth}
\centering
\includegraphics[width=1.8in]{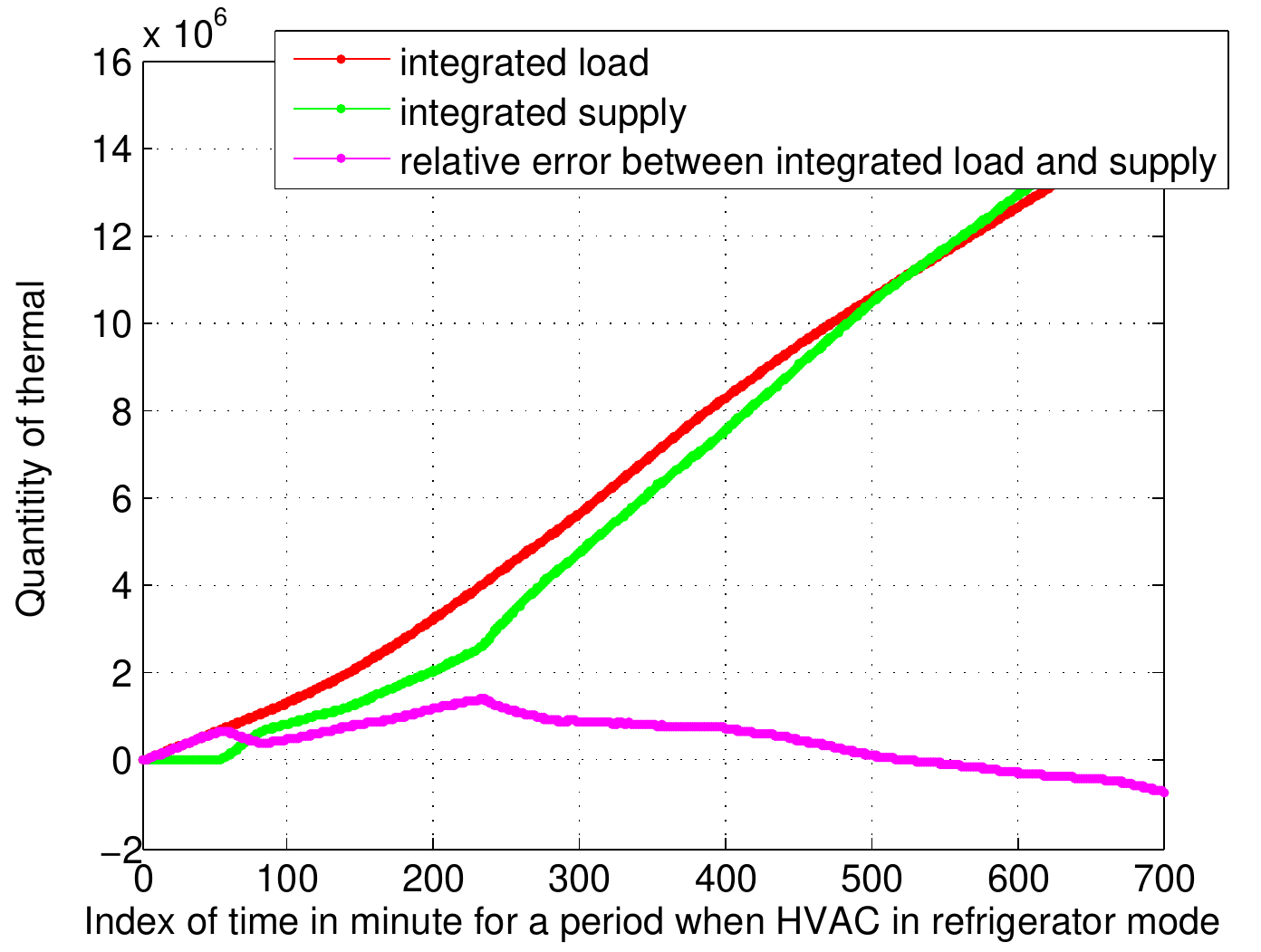}
\end{minipage}%
\caption{Derived load signatures Vs. Variation Signatures of Supply Vs. the relative error between integrated load and integrated supply}
\label{signature}
\end{figure*}
\subsection{Identify Load Signature by Linear Regression}
Let's consider the joint impacts of load and supply to the indoor temperature.  Without losing of generality, let's consider the case when the HVAC is working in the refrigerator mode, by substituting (\ref{load}) and (\ref{supply1}) into (\ref{equal}), we have:
\begin{equation}
\left[ {\begin{array}{*{20}{c}}
  {n(t)({T_p} - T(t))} \\ 
  {{T_o}(t) - T(t)} \\ 
  {V_{cool}^w(T_{in}^w(t) - T_{out}^w(t))} 
\end{array}} \right]^{T}\left[ {\begin{array}{*{20}{c}}
  {{c_p}} \\ 
  {{\alpha}} \\ 
  {{-\beta _{ac}}} 
\end{array}} \right] = c{M_z}\Delta (t)
\label{array1}
\end{equation} 
where $\alpha=cM_{air}+\frac{1}{R_{eq}}$ is the coefficients of $T_{o}(t)-T(t)$ in the load model, which is modeled as one unknown coefficient. We can rewrite (\ref{array1}) as $\mathbf{A}(t)\mathbf{\theta}=\mathbf{B}(t)$.  Then by sensor measurements and HVAC states from 1 to t, we can set up an overdetermined observation matrix $\mathbf{A}_{1:t}=[\mathbf{A}(1), \mathbf{A}(2), \cdots, \mathbf{A}(t)]^{T}$, and an observation vector $\mathbf{B}_{1:t}=[\mathbf{B}(1), \mathbf{B}(2), \cdots, \mathbf{B}(t)]^{T}$. Then the problem of identifying the load signature is to identify the vector $\mathbf{\theta}$ by solving $\mathbf{A}_{1:t}\theta=\mathbf{B}_{1:t}$, with the constraints that  $c_{p}, \alpha, \beta_{ac}$ are nonnegative.

\section{Techniques to Solve the Regression Model by Real Data}
We used real data collected from HaiDianHuangZhuang Station to calculate the model parameters in (\ref{array1}) and to investigate the signatures of the loads. 
\subsection{Calculate Coefficients by Real Data}
Data collected from HaiDianHuangZhuang station from a timespan of Aug 21th, 2013 to Aug 23th, 2013 was selected to solve the linear regression model. The dataset provides real-time $T(t)$, $T_{ac}(t)$, $T_{in}^{w}(t)$, $T_{out}^{w}(t)$, $V_{cool}^{w}$, and $E_{v}$, which are in one-minute resolution.  In addition, passenger flows are acquired by the ticket checking system in per-hour resolution. We estimated the per-minute resolution passenger amount by linear  interpolations. Based on these data, the observation matrix $\mathbf{A}_{1:t}$ is constructed and the vector $\mathbf{B}_{1:t}$ are constructed. Note that the volume of air $M_{z}$ in the subway station  is inferred by the geometrical data of the station. 

Since the coefficients are required to be nonnegative, directly applying  the least square estimation is inefficient. We propose a search algorithm to solve this constrained optimization problem:
\begin{equation}
\begin{gathered}
  \theta  = \mathop {\arg \min }\limits_{\left[ {{c_p},\alpha ,{\beta _{ac}}} \right]} \frac{{\sum\limits_{i = 1}^t {\left| {{\mathbf{A}_{i,1}}{c_p} + {\mathbf{A}_{i,2}}\alpha  - {\mathbf{A}_{i,3}}{\beta _{ac}} - {\mathbf{B}_i}} \right|} }}{{\sum\limits_{i = 1}^t {\left( {{\mathbf{A}_{i,1}}{c_p} + {\mathbf{A}_{i,2}}\alpha } \right)} }} \hfill \\
  {\text{subject to: }}{c_p} > 0,\alpha  > 0,{\beta _{ac}} \ge  0 \hfill \\ 
\end{gathered} 
\label{optproblem}
\end{equation}

$\mathbf{A}_{i,j}$ is the item in $i$th column and $j$th row in the matrix $\mathbf{A}_{1:t}$. Note that we divide the accumulated absolute difference of the loads and the supplies by the accumulated loads, which is to find the coefficient vector that can provide the minimum relative difference between the load vector and the supply vector.  Otherwise, smaller parameters providing smaller absolute error tend to be voted for the lacking of normalization. The search algorithm searches all combinations of $[c_{p}, \alpha]$ for $c_{p}<1000$ and $\alpha<10000$. For each combination of $c_{p}$ and $\alpha$, $\beta_{ac}$ that provides the minimum relative error is calculated. The parameter set $[c_{p}^{*}, \alpha^{*}, -\beta_{ac}^{*}]$ which provides the overall minimum relative error is chosen as the optimal solution of problem (\ref{optproblem}). For the number of coefficients is limited, the computing complexity of the algorithm is tolerable. 
\subsection{The Load Signatures}
Another difficulty to solve (\ref{optproblem}) is that we found the vector $B_{1:t}$ is highly zigzagging over time, which is due to the noises of the measurements of the temperature sensors, i.e.,  the difference of indoor temperature of successive time cannot be accurately measured  because of the accuracy limitation of sensors. To overcome this noise issue, we proposed to further calculate the coefficients by minimizing the differences of the \emph{integrated loads} and the \emph{integrated supply}.  We define the difference between the integrated load and the integrated supply by $\mathbf{C}(T)=\sum_{t=1}^{T}\mathbf{A}(t)$; the integrated indoor thermal variation is defined by $\mathbf{D}(T)=\sum_{t=1}^{T}\mathbf{B}(t)$. Then we solve (\ref{optproblem}) by replacing $\mathbf{A}_{i,j}, \mathbf{B}_{i}$ by $\mathbf{C}_{i,j}, \mathbf{D}_{i}$. This method gives us robust estimation of the coefficients which can tolerates the sensor noises.  For the particular dataset of August 23, 2013 of HaidianHuangZhuang station, we calculated the optimal parameter set  $\theta$ as $[83, 53703,  -1290071]^{T}$. When varying the scope of the data, we found the solution vary within tolerable range of errors. 

By substituting the calculated coefficients into the load model, the derived load signature was plotted in Fig.\ref{signature}a). It shows that \emph{the loads from the outdoor temperature take the major portion, while the thermal loads introduced by the passengers take a small portion}. The real-time supplies calculated by the supply model are plotted in  Fig.\ref{signature}b). We can see the variation of supply has similar pattern as the load. The relative error between the integrated load and integrated supply is plotted in Fig.\ref{signature}c), which it is relative small by calculating using the optimally derived parameters.  It indicates that the searching algorithm has provided a rather confident estimation to the load signatures. 
 
\section{Conclusion and Discussion}
This paper investigated the load signatures of HVAC system in subway station based on real data collected from subway station. By extensive sensor data collected from environments and the HVAC system, we proposed a linear regression to model to describe the impacts of loads and the cooling supply to the indoor temperature. We then present a search algorithm to identify the model coefficients by minimizing the integrated differences between load and supply, which can tolerate the noises of sensor measurements.  Experiment results on real dataset show the proposed method can provide rather confident load signature which highly coincides with the real-time supply measurements. Since the load signature provide important knowledge for the energy efficient control, we will study the optimal control strategies in our future work.

\bibliographystyle{abbrv}
\bibliography{ref}
%

\end{document}